\documentclass[twocolumn,aps,amssymb]{revtex4}
\usepackage{latexsym}
\usepackage{graphicx}

\begin{document}
\keywords{Electron glass, ageing, disordered insulators.}



\title{Ageing in granular aluminium insulating thin films}



\author{Julien Delahaye and Thierry Grenet}
\address{Institut N\'eel, CNRS-UJF, BP 166, 38042 Grenoble, France}

\begin{abstract}
  We present a new set of electrical field effect measurements on granular aluminium insulating thin films. We have explored how the conductance relaxations induced by gate voltage changes depend on the age of the system, namely the time elapsed since its quench at low temperature. A clear age dependence of the relaxations is seen, qualitatively similar to ageing effects seen in other well studied glassy systems such as spin glasses or polymers. We explain how our results differ from the previous ones obtained with different protocols in indium oxide and granular aluminium thin films. Our experimental findings bring new information on the dynamics of the system and put new constraints on the theoretical models that may explain slow conductance relaxations in disordered insulators.
\end{abstract}

\maketitle                   






\section{Introduction}

Among the electronic properties of disordered insulators, the slow relaxation of their
electrical conductance is quite outstanding. This relaxation found up to now in a limited
number of systems has been studied in many details in indium oxide (InOx) \cite{OvadyahuPRB91,OvadyahuPRB02}
and granular aluminium (Al) \cite{GrenetEPJB07} thin films. The main features are the following.
When the samples are quenched to liquid helium temperature, their conductance subsequently decreases like a logarithm of time for weeks without any tendency
to saturation. Moreover, it was shown in MOSFET devices that a change of the gate
voltage induces a fast conductance increase followed by a new slow and logarithmic
decrease. Fast gate voltage sweeps $G(V_g)$ performed after a long stay under a fixed gate voltage $V_{g1}$ display a dip centred on $V_{g1}$, the so called anomalous field effect \cite{OvadyahuPRB91}. The amplitude of the conductance decrease under a fixed gate voltage and the corresponding growth of the $G(V_g)$ dip depend on the temperature and on the resistance per square of the samples, a few \% being typical for measurable samples.

These slow relaxation phenomena have aroused a broad interest since they may be the first direct experimental signature of the Coulomb glass, a glassy state resulting from the coexistence of disorder and electron-electron interactions and predicted more then 20 years ago \cite{DaviesPRL82}. But in spite of important theoretical developments \cite{MullerPRL04,MullerPRB05,MalikPRB07,SurerPRL09}, their detailed understanding is still unachieved.

In this article, we present new experiments in which we study the age dependence of the gate voltage induced conductance relaxations. According to the usual definition in other glasses, we call the age of the samples the time elapsed since their quench at low temperature. In order to clarify what is new compared with previous measurements, we first review some older protocols which were also claimed to demonstrate ageing effects. We then expose our new ageing protocols and what results we indeed get from them. And finally, we discuss the possible physical meaning of our findings.

\section{Experimental}

The samples used in this study are granular aluminium thin films, $20~nm$ thick, made by e-gun evaporation of Al under a partial pressure of oxygen. The substrates are either highly doped silicon wafers or sapphire single crystals covered with an Al layer. A gate insulator $100nm$ thick (silicon dioxide or alumina) enables us to measure the conductance of the films as a function of a gate voltage. All the films are insulating, with an exponential-like divergence of the resistance at low temperature. The electronic transport in the film is believed to occur through electron tunneling between nanometric Al grains. Resistances per square range between $10G\Omega$  and $1M\Omega$ at liquid helium temperature. Fuller details of samples preparation, characterization and measurements techniques are given elsewhere \cite{GrenetEPJB07}. All the measurements presented here were performed at $4.2K$.

\section{Previous protocols}

Different protocols have been built on in order to explore the response of InOx and granular Al thin films to gate voltage changes. But they are generally performed a long time after the quench of the samples. By long, we mean that the samples are kept under a fixed gate voltage $V_{g1}$ until the conductance relaxation of the system is negligible over the time scale of the experiments (quasi-equilibrium state). This first waiting time ($t_{w1}$) at $V_{g1}$ that we will call later "the age of the system", is typically of a few days or more.

\begin{figure}[h]
\begin{center}
\includegraphics[height=7cm]{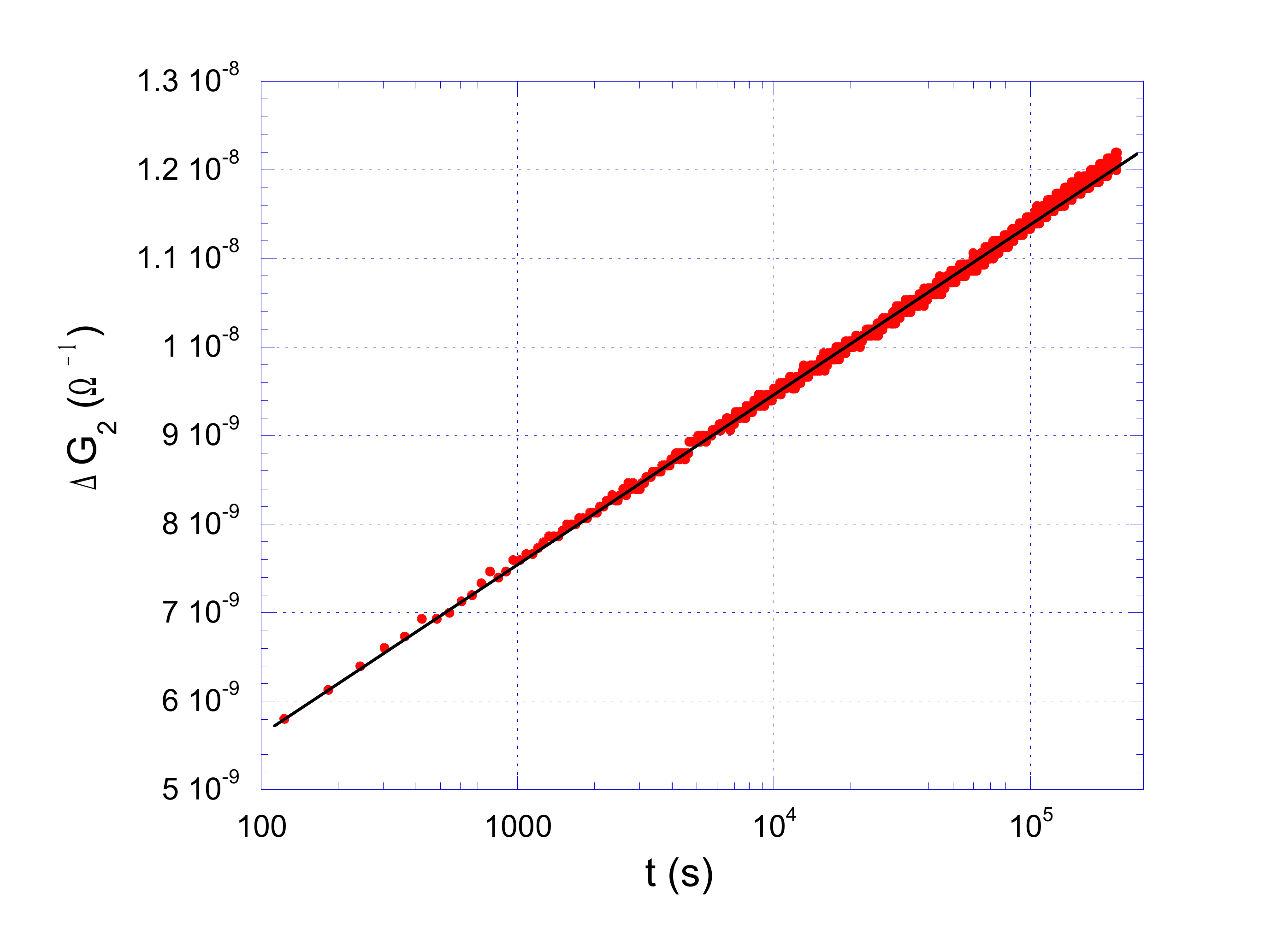}
\end{center}
\caption{Amplitude of a new conductance dip a time t after the gate voltage was changed from $V_{g1} = 0V$  to $V_{g2} = -30V$. The sample was kept 5 days at $4.2K$ under $V_{g1}$ before changing $V_g$ ($t_{w1}= 5days$). $R_\Box = 30M\Omega$. }
\label{fig:1}
\end{figure}

The simplest protocol (protocol 1) consists in changing after a waiting time $t_{w1}$ the gate voltage to a new value $V_{g2}$. The conductance increases rapidly to an off-equilibrium value and then starts to decrease as a logarithm of time. Fast $G(V_g)$ sweeps around $V_{g2}$ at constant time intervals reveal a corresponding $\log(t)$ growth of the conductance dip centred on $V_{g2}$ (dip 2). This is illustrated in Fig.~\ref{fig:1}, where the $\log (t)$ increase of dip 2 amplitude $\Delta G_2$ is observed over 3 orders of magnitude in time (deviations from a perfect $\log(t)$ are indeed visible at the longest times, as will be discussed below).

\begin{figure}[h]
\begin{center}
\includegraphics[height=7cm]{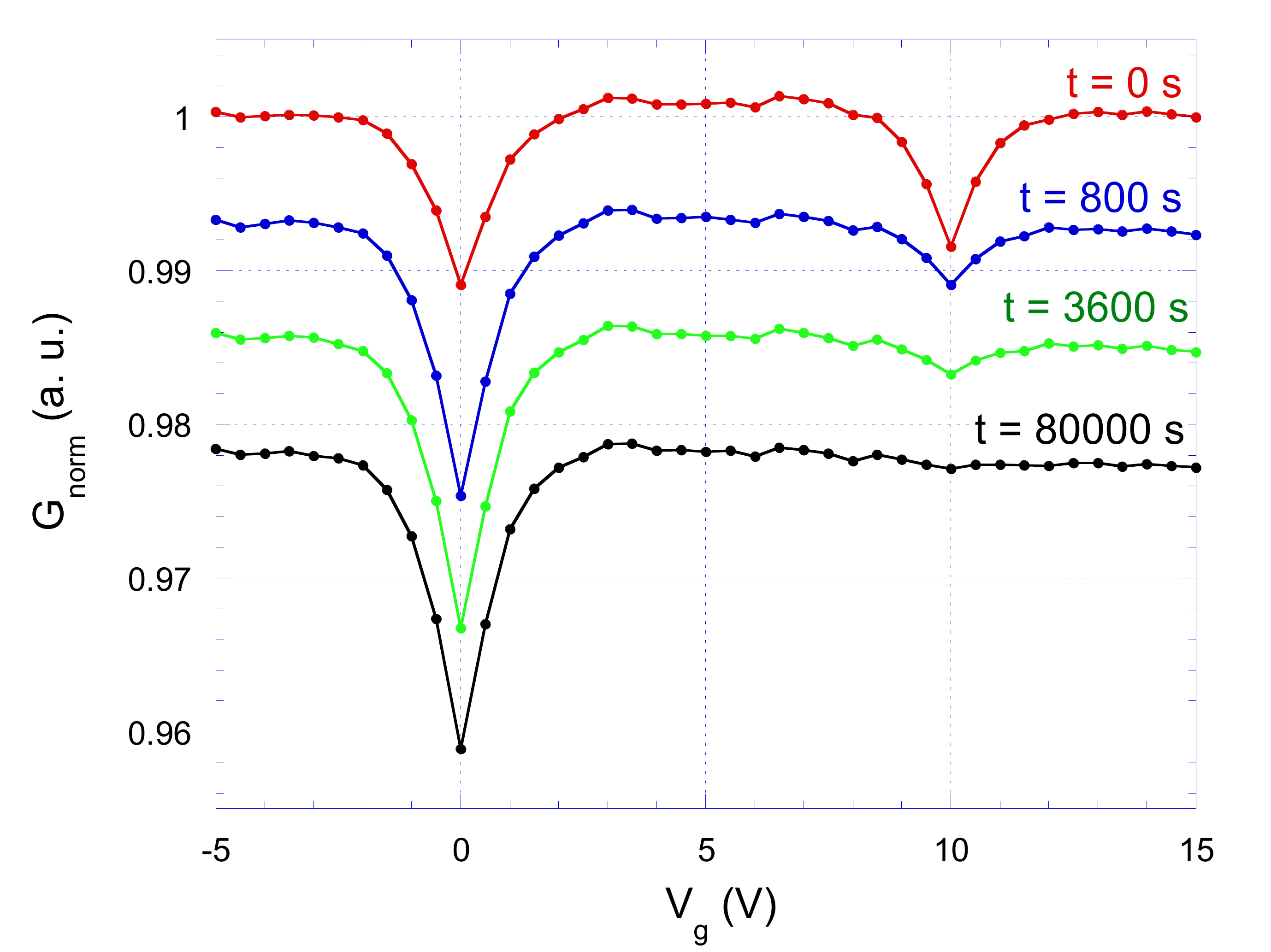}
\end{center}
\caption{$G(V_g)$ curves during a two dips measurement (see text for details). The sample was maintained $20h$ under $V_{g1}=0V$ and then $1h$  under $V_{g2}=10V$. The first $G(V_g)$ scan was taken just before $V_g$  was switched back to $V_{g1}$ ($t=0$). $G(V_g)$ scans $80s$ long  are taken every $400s$ to follow the erasure of the new dip at $10V$ and the recovery of the old one at $0V$. Small mesoscopic reproducible conductance fluctuations are superimposed on the conductance dips relaxations (see \cite{DelahayeEPJB08}  for more details). $R_\Box = 15M\Omega$. The $V_{g2}$ dip is almost completely erased after $20h$. The curves have been shifted for clarity.}
\label{fig:2}
\end{figure}

A more complex protocol (protocol 2) called the 2 dips measurement was introduced by Zvi Ovadyahu et al in InOx films \cite{OvadyahuPRL00}. In this protocol, the gate voltage is changed after $t_{w1}$ to a new value $V_{g2}$, kept there during a time $t_{w2}$ and then switched back to $V_{g1}$ (time $t=0$ in Fig.~\ref{fig:2}). Just before $V_g$ is switched back to $V_{g1}$, a new dip has formed at $V_{g2}$ while the old one at $V_{g1}$ has been partly erased (see the upper curve of Fig.~\ref{fig:2}).

\begin{figure}[h]
\includegraphics[width=68mm]{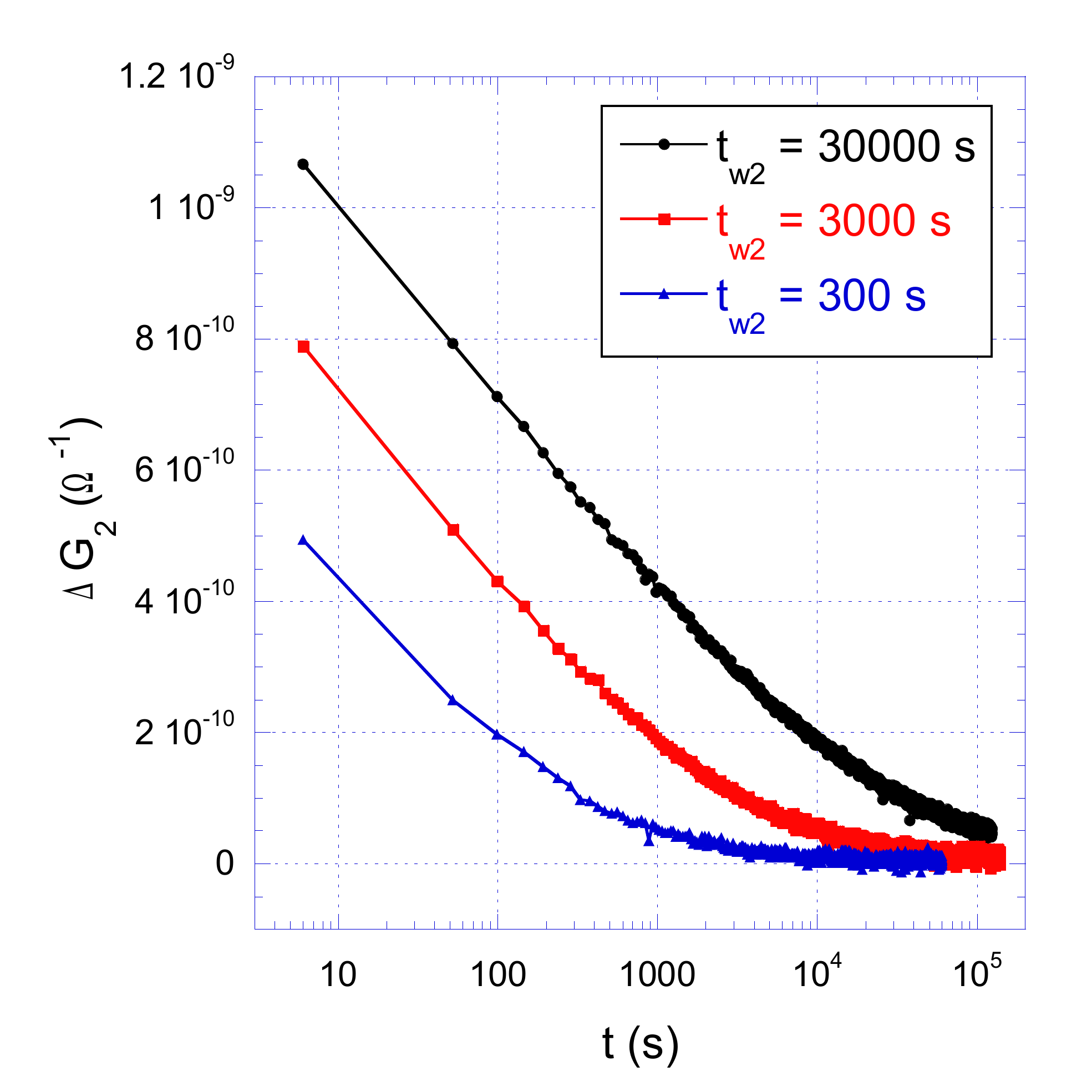}~a)
\hfil
\includegraphics[width=68mm]{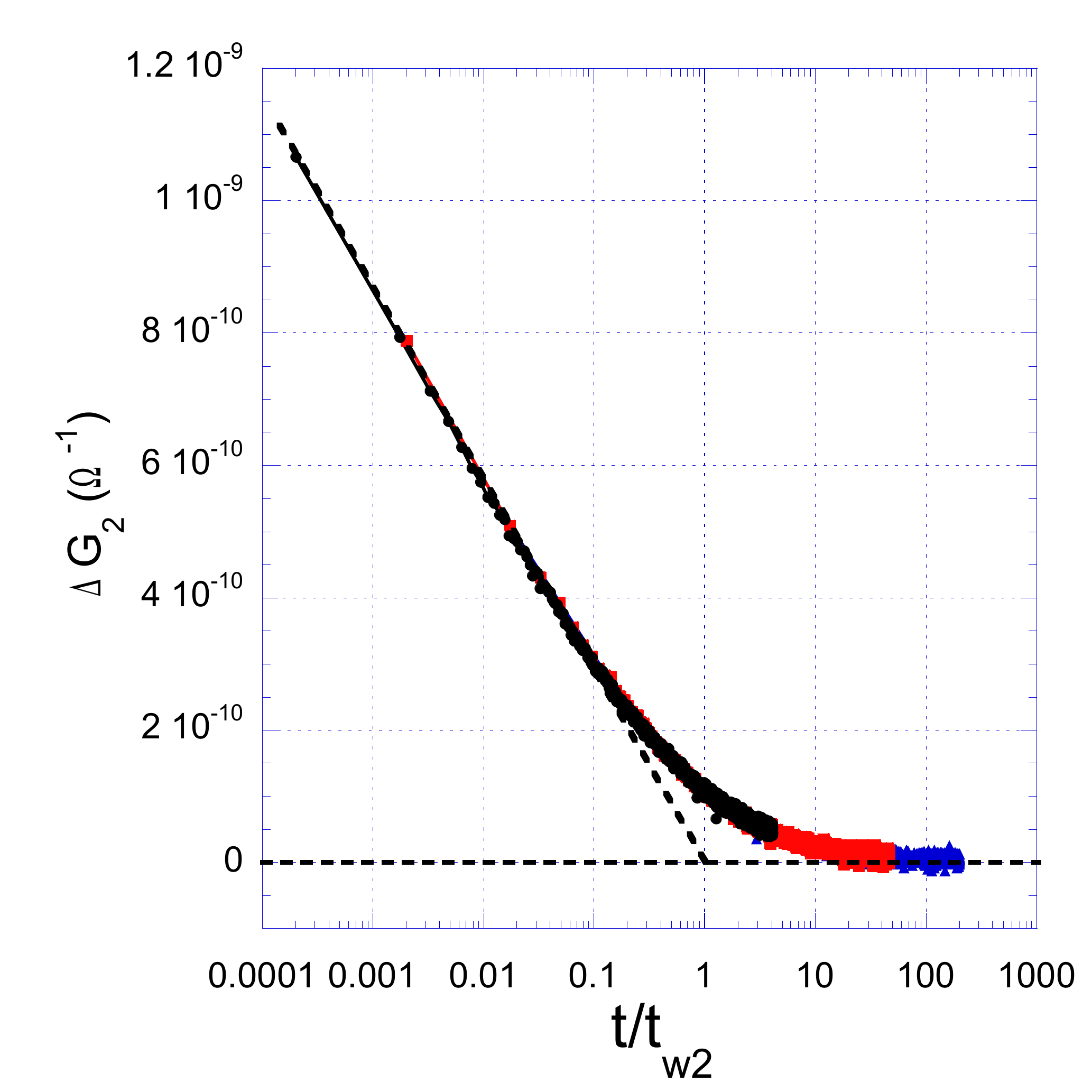}~b)
\caption{Erasure of the conductance dip 2 dug during $t_{w2}$ at $V_{g2}$ after $V_g$ is switched back to $V_{g1}$ (2 dips experiment). (\textbf{a}) Amplitude of the dip as a function of time $t$. (\textbf{b}) Amplitude of the dip as a function of the reduced time $t/t_{w2}$. The extrapolation of the short times logarithmic part is shown with intersect of the doted line (see text for details). $R_\Box=480~M\Omega$.}
\label{fig:3}
\end{figure}

By repeating fast $G(V_g)$ scans, we can then follow as a function of time how the conductance dip at $V_{g1}$ goes back to its quasi-equilibrium value (recovery of dip 1) or look how the dip at $V_{g2}$ is erased (erasure of dip 2). This is illustrated in Fig.~\ref{fig:2} and \ref{fig:3}. Both lead to the same result, namely the longer $t_{w2}$ is, the longer it takes for the system to come back to its quasi-equilibrium state. More precisely, the relaxation curves to quasi-equilibrium (see Fig.~\ref{fig:3}) are well described by a unique function of $t/t_{w2}$ (full scaling). It was this $t_{w2}$ dependence of the relaxation which was called ageing and even simple ageing as long as the $t/t_{w2}$ scaling is obeyed \cite{OvadyahuPRL00,GrenetEPJB07}. It is also noteworthy that the extrapolation of the logarithmic part of the curves visible at short times indicates that the quasi-equilibrium state is recovered after a characteristic time $t_{w2}$ (extrapolated amplitude of dip 2 equal to $0$).

It was shown in \cite{GrenetEPJB07} that the full scaling obtained with the protocol 2 can be well understood simply by assuming that the conductance relaxations result from the additive contributions of independent and reversible degrees of freedom having a $1/\tau$ distribution of relaxation times. Similar ideas were recently develop within the Coulomb glass picture frame \cite{AmirPRB08,AmirCondmat09}. The full scaling of Fig.~\ref{fig:3} may thus be understood without assuming any change in the dynamics of the system, contrary to the usual definition of ageing in glasses \cite{BiroliJSM05}.

\section{New protocols}

In a new set of measurement, we have tested how the conductance relaxation depends on the age $t_{w1}$ of the system. More precisely, we have measured how the results of protocols 1 and 2 are modified when the time $t_{w1}$ is of the order and smaller than $t_{w2}$ and the time scale of the measurements. This procedure is very similar to pioneer ageing experiments done in polymers \cite{StruikBook78}.

\begin{figure}[h]
\begin{center}
\includegraphics[height=7cm]{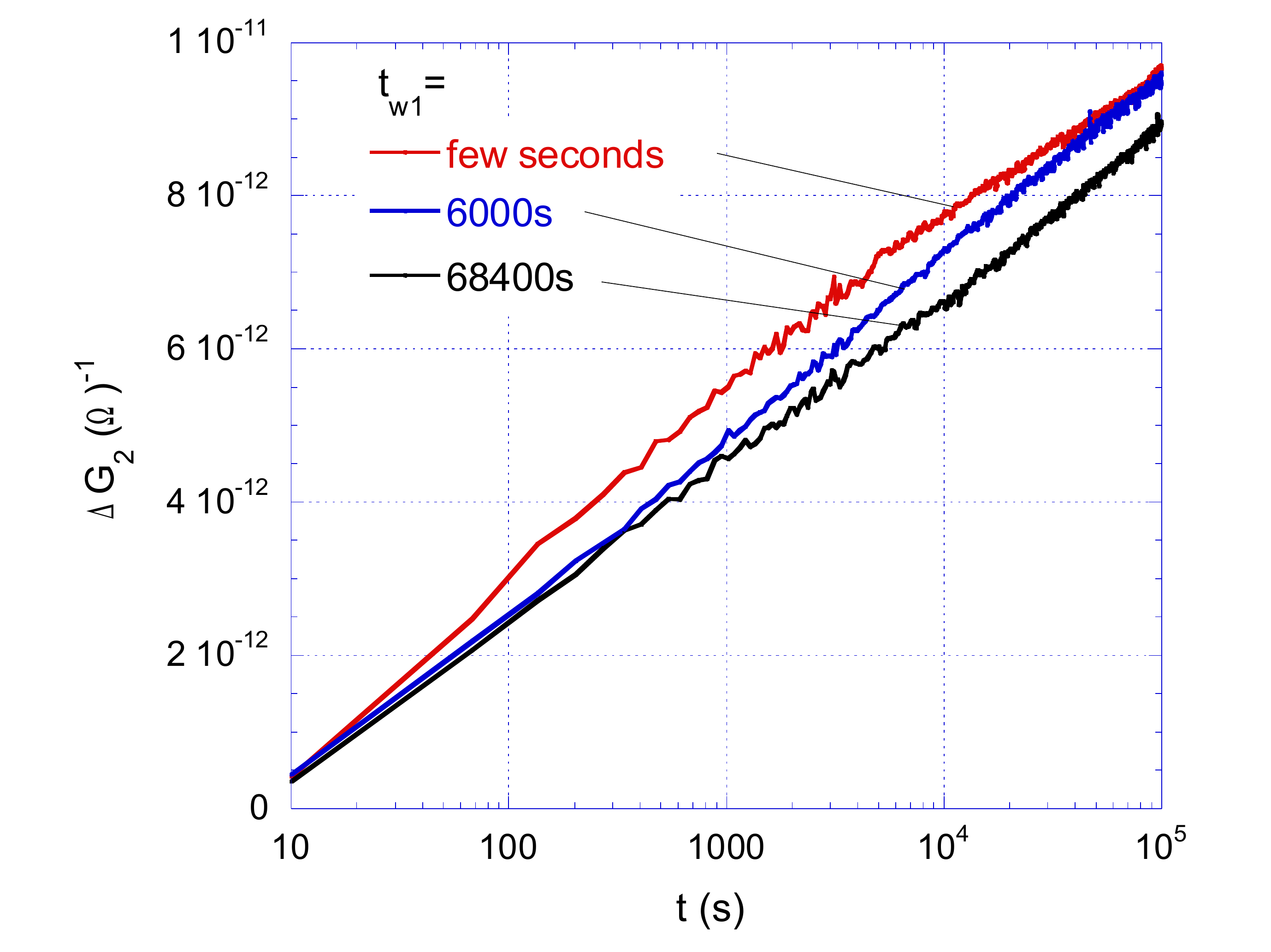}
\end{center}
\caption{Growth of the conductance dip amplitude after a gate voltage change for different $t_{w1}$s. $R_\Box=5G\Omega$. The results should be compared with that of Fig.~\ref{fig:1}.}
\label{fig:4}
\end{figure}

\begin{figure}[h]
\begin{center}
\includegraphics[height=7cm]{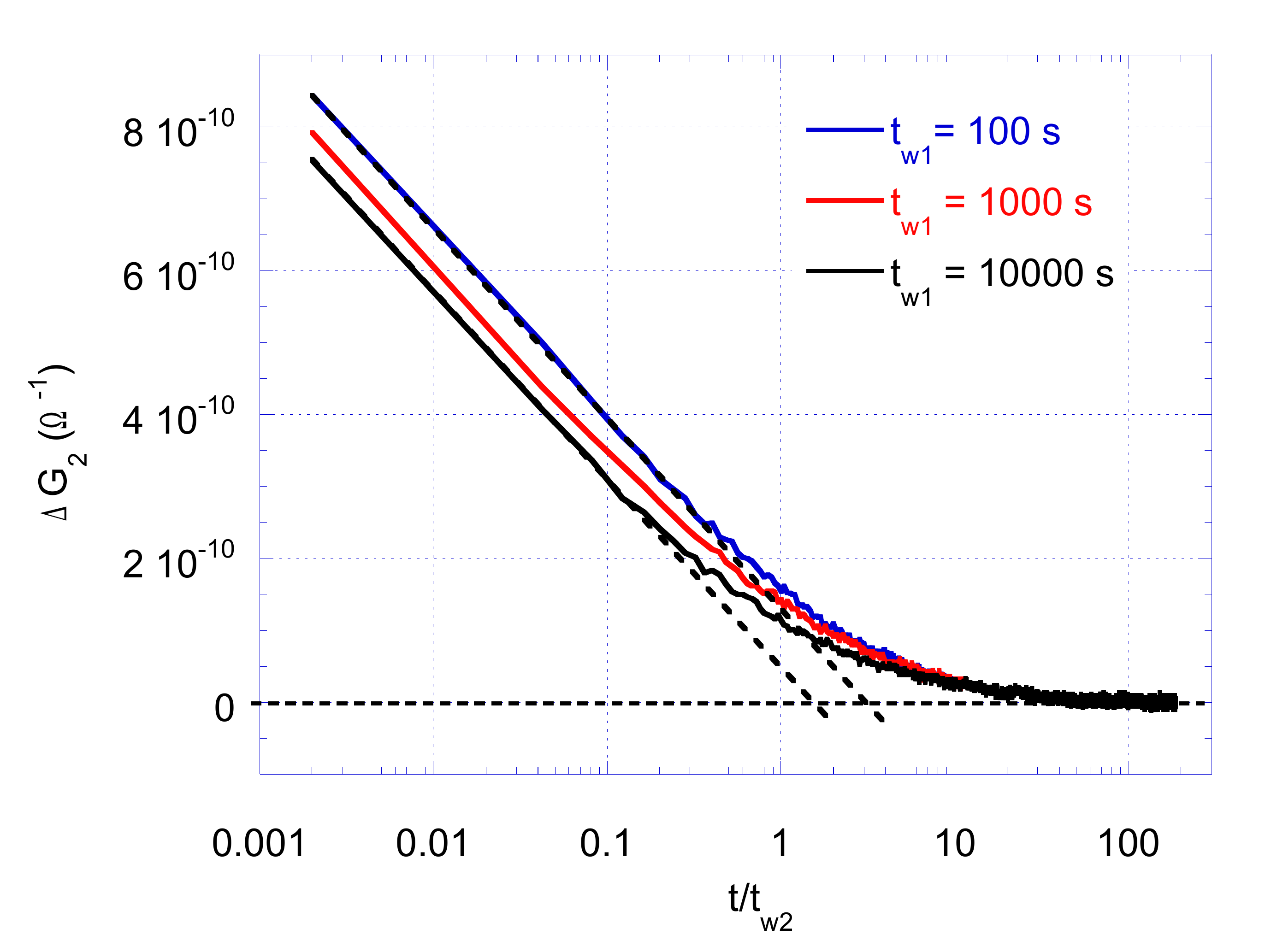}
\end{center}
\caption{: Erasure of the conductance dip 2 in a 2 dips experiment for different ages $t_{w1}$s (see text for details). The time $t_{w2}$ is fixed to $1000s$. The erasure characteristic times are given by the intersect of the doted lines. $R_\Box=480M\Omega$.}
\label{fig:5}
\end{figure}

By repeating the protocol 1 for different $t_{w1}$s, we now clearly see (Fig.~\ref{fig:4}) that the increase of a new conductance dip does not follow a pure $\log(t)$ dependence. The  $G(t)$ curves are s-shape, with a faster dip amplitude increase around $t = t_{w1}$.
By repeating the protocol 2 for different $t_{w1}$s, we have found two interesting features. First, the larger $t_{w1}$ is, the longer it takes to erase the conductance dip at $V_{g2}$ for a fixed $t_{w2}$ (Fig.~\ref{fig:5}). The extrapolation of the logarithmic part of the relaxation gives a time larger than $t_{w2}$ as the typical erasure time, $t_{w2}$ being the $t_{w1}$ infinite limit. Thus the erasure "master function" is modified for "young" samples and the form shown in Fig.~\ref{fig:3} is only the limiting case of very "old" samples ($t_{w1}\gg t_{w2}$). However, even for $t_{w1}$ close to $0$, the full scaling as a function of $t/ t_{w2}$ is conserved within our experimental accuracy (not shown here).

\section{Discussion}

The results above are in qualitative agreement with what is found in other glassy systems \cite{StruikBook78,VincentLN08}. The s-shape visible in Fig.~\ref{fig:4} was also found in spin glasses \cite{VincentLN08} where it was interpreted as a maximum in the $log$ distribution of relaxation times. A consequence of Fig.~\ref{fig:4} is that the longer one waits after the quench, the longer it will take to dig a new conductance dip of a given amplitude, in qualitative agreement with a hardening of the system with its age. The extrapolation of the logarithm relaxation part to times larger than $t_{w2}$ visible in Fig.~\ref{fig:5} indicates a symmetry breaking between the digging of the dip at $V_{g2}$ and its erasure under $V_{g1}$. Such symmetry breaking is also in agreement with a slower response as the system ages: it takes a longer time to erase the dip after $t_{w1} + t_{w2}$ than it took to dig it between $t_{w1}$ and $t_{w1} + t_{w2}$. The fact that the $t / t_{w2}$ scaling is conserved is more puzzling and deserves further theoretical analysis.

Clearly and contrary to previous results, our ageing effects cannot anymore be described by the linear contribution of independent degrees of freedom relaxing back and forth with the same relaxation times. More involved models, e.g. with relaxation time distributions differing from pure $1/\tau$ and depending on the system's age may be needed. Whereas it is possible to propose an interpretation in agreement with the Coulomb glass theoretical picture still needs to be clarified.
Finally, we would like to point out that a similar symmetry breaking was also observed in InOx thin films \cite{OvadyahuPRB02,OvadyahuPRB03,OvadyahuPRL04}. Starting from the pseudo-equilibrium state ($t_{w1} \gg t_{w2}$), the conductance relaxation at $V_{g1}$ was followed after large $V_g$ or bias voltage changes maintained during $t_{w2}$. As long as these changes are not too large (small loss of memory) the symmetry is broken (the extrapolation of the logarithmic part of the relaxation gives characteristic times larger than $t_{w2}$) but the full scaling is conserved. This is fully consistent with our results and simply indicates that such modified erasure functions are obtained for "young" systems and that we have different ways to rejuvenated the systems.

\section{Conclusion}

In conclusion, we have observed for the first time that the conductance relaxation in granular aluminium thin films depends on the age of the samples, i.e. the time elapsed since their quench at low temperature. These ageing effects are in qualitative agreement with what was found in other glassy systems, such as polymers or spin glasses. It would now be interesting to see if they are also present in InOx films and if they can be explained within the frame of the Coulomb glass. Our observations also open new issues, such as memory and rejuvenation experiments as a function of temperature.

\section{acknowledgement}
  We acknowledge financial support from the "Jeunes Chercheurs" program of the French National Research Agency ANR (contract $N^o$ JC05-44044).

\end{document}